\documentclass[aps,twocolumn,epsfig,graphics,showpacs,floatfix,mathbbm]{revtex4}

\usepackage{amsmath,amsfonts,amssymb,graphics,graphicx,epsfig,color,times,bbm}

\begin{document}
\bibliographystyle{apsrev}
\newcommand{\proofend}{\hfill\fbox\\\medskip }

\title{Towards mechanical entanglement in
nano-electromechanical devices}

\author{J.\ Eisert$^{1,2}$,  M.B.\ Plenio$^{2}$,  S.\ Bose$^{3,4}$, and J.\ Hartley$^{2}$ }

\affiliation{
1 Institut f{\"u}r Physik, Universit{\"a}t Potsdam,
Am Neuen Palais 10, D-14469 Potsdam, Germany\\
2 QOLS, Blackett Laboratory, Imperial College London,
Prince Consort Road, London SW7 2BW, UK\\
3 Department of Physics and Astronomy, University College London,
Gower Street, London, WC1E 6BT, UK\\
4 Institute for Quantum Information, California Institute of
Technology, Pasadena, CA 91125, USA}

\date{\today}

\begin{abstract}

We study arrays of mechanical oscillators in the quantum domain
and demonstrate how the motions of  distant oscillators can be
entangled without the need for control of individual oscillators
and without a direct interaction between them. These oscillators
are thought of as being members of an array of
nano-electromechanical resonators with a voltage being applicable
between neighboring resonators. Sudden  non-adiabatic switching of
the interaction results in a squeezing of the states of the
mechanical oscillators, leading to an entanglement transport in
chains of mechanical oscillators. We discuss spatial dimensions,
$Q$-factors,  temperatures and decoherence sources in some detail,
and find a distinct robustness of the entanglement in the
canonical coordinates in such a scheme. We also briefly discuss
the challenging aspect of detection of the generated entanglement.

\end{abstract}

\pacs{03.67.-a, 07.10.Cm, 03.65.Yz}

\maketitle

In 1959 Richard Feynman suggested in a famous talk that it appears
to be a fruitful enterprise to think about manipulating and
controlling mechanical devices at a very small scale. Since then,
the study of micro-electromechanical (MEMS) and even
nano-electromechanical systems (NEMS) has developed into a mature
field of research
\cite{Roukes,Review,Measurement,Measurement2,Schwab}. Mechanical
oscillators with spatial dimensions of a few nanometers and very
 high frequencies can now be routinely manufactured.
Applications of such NEMS range from mechanically-detected
magnetic resonance imaging, sensing of biochemical systems, and
ultrasensitive probing of thermal transport. Indeed, the NEMS
devices that are presently manufactured in experimental studies
are close to or already on the verge of the quantum limit
\cite{Roukes,Review,Measurement,Measurement2,Schwab}. While first
quantum effects are already being observed and studied, it  is
interesting to see to what extent it is feasible to prepare
nano-scale mechanical oscillators in states where the quantum
nature becomes most manifest: in states that are genuinely
entangled in the canonical coordinates of position and momentum.
This can be interesting for a variety of reasons. Firstly, it
provides another stepping stone towards quantum state control and
quantum information processing in mechanical systems. This is
particularly fascinating as these systems are macroscopic
consisting of many million atoms. They would therefore also permit
the exploration of the limiting region between the quantum and the
classical world. This might be facilitated by another application
of entanglement namely its use to enhance quantum measurement
schemes where entangled states represent  a very sensitive probe.

The key question that will be addressed in this letter is how it
is possible to entangle mechanical oscillators well separated in
space,  without the need for making them  interact directly and
with a minimum need for individual local control which is
difficult to achieve at the nano-level. This  will be accomplished
by triggering squeezing and entanglement by a global non-adiabatic
change of the interaction strength in a linear array of
oscillators, but without individually addressing any of the
oscillators of the array.
% As a consequence of this sudden non-adiabatic change one
% encounters local squeezing of the individual oscillators
% which is then transformed into entanglement via the system
% dynamics. At the same time this entanglement is transported
% through the chain in the sense that there is a finite time
% after changing the interaction parameter that remotely located
% oscillators become entangled.
 In this way, one can achieve long-range entanglement that
will persist over length scales that are much larger than the
typical entanglement length for the ground state of the system
\cite{Chain}. The physics underlying this approach, especially the
non-adiabaticity requirement, will be discussed in more detail
lateron. Several schemes to probe quantum coherence of mechanical
resonators in different setups and situations have been proposed
so far \cite{blencowe02,mancini}. Notably, while the earlier
proposal of entangling macroscopic oscillators \cite{mancini}
entangles two adjacent oscillators in the context of a different
physical setup, our scheme allows, without the need for individual
local control, for entanglement in the canonical coordinates
between non-adjacent (and possibly distant) microscopic
oscillators by entanglement transport in a chain.

 The setup that we will consider is an array of double-clamped
coupled nano-mechanical oscillators as has been experimentally
studied in the micro-mechanical realm in Ref.\ \cite{Buks}.
%We consider an array of doubly clamped beams, each of which being a
%mechanical oscillator.
We assume that the beams are arranged in such a manner that
between adjacent oscillators a controlled and  tunable interaction
can be introduced. In Ref.\ \cite{Buks} this is experimentally
achieved  by applying a voltage between adjacent beams made from
gold fabricated on a semiconductor membrane that are ordered
alternatingly. This induces to a good approximation a nearest
neighbor interaction that can be controlled in strength. The
oscillators are assumed to be cooled to temperatures such that
 $kT/\hbar \omega\ll 1$ with  $\omega$ being the
fundamental frequency of the oscillators, such that the array is
operated deeply in the quantum regime.
Before we discuss the time and energy scales that would be
required to achieve this regime, we will exemplify the mechanism,
without taking sources of error and decoherence mechanisms into
account, as we will discuss these in some detail later.
We start with the  Hamiltonian of $N$ quantum oscillators  of mass
$m$ and eigenfrequency $\omega$ ordered on a one-dimensional
lattice, with nearest-neighbor interaction of strength $c$.
Setting $\hbar=1$ and using the $q_{k}= q_k'\sqrt{m\omega}$, and
$p_{k}= p_k' /\sqrt{m\omega}$, where $q_k' $ and $p_{k}'$ are  the
canonical
% the canonical coordinates reflecting
position and momentum of the oscillators  we find
\begin{equation}\nonumber
    H=\frac{\omega }{2}\sum_{k=1}^N
    \Bigl(
    p_k^2 + q_k^2 (1+2 c) - 2c q_k q_{k+1}
    \Bigr),
\end{equation}
%
%where $N$ is the number of oscillators,
For the moment, we assume for simplicity periodic boundary
conditions, i.e., $q_{N+1}=q_1$, but this requirement will be
relaxed later, and set $\omega=1$, as in this ideal treatment this
merely corresponds  to a rescaling of the time scale. The normal
coordinates are related to the previous ones by a discrete Fourier
transform,
   $ q_k = (1/\sqrt{N})
    \sum_{l=1}^N e^{ 2\pi i k l / N} Q_l$,
    $p_k =  (1 / \sqrt{N})
    \sum_{l=1}^N e^{ -  2\pi i k l /N } P_l$.
%
%\begin{eqnarray*}
%    q_k &=& \frac{1}{\sqrt{N}}
%    \sum_{l=1}^N e^{\frac{2\pi i k l}{N}} Q_l,\,\,
%    p_k =  \frac{1}{\sqrt{N}}
%    \sum_{l=1}^N e^{ - \frac{2\pi i k l}{N}} P_l.
%\end{eqnarray*}
In these normal  coordinates, satisfying $Q_k=Q^\dagger_{N-k}$ and
$P_k=P^\dagger_{N-k}$, the Hamiltonian can be written in the form
\begin{equation}\nonumber
    H=\frac{1}{2}\sum_{k=1}^N
    \left(P_k P_k^\dagger + (1+ 4 c \sin^2 (\pi k/N) )^{1/2} Q_k Q_k^\dagger
    \right),
\end{equation}
 %where $\omega_k=(1+ 4 c \sin^2 ( \pi k/N) )^{1/2}$. Introducing
annihilation and creation operators,
%In terms of the annihilation and creation operators, $a_k$ and $a_k^\dagger$,
%defined as
%\begin{equation}
%   a_k = ( \omega_k Q_k + i P_k^\dagger)/\sqrt{2 \omega_k},
%\end{equation}
%one ... introduce the Heisenberg,
and expressing the time dependent operators $Q_k(t)$ and $P_k(t)$
in terms of these operators, one arrives at the Heisenberg
equations of motion
for the original canonical coordinates
\begin{eqnarray*}
    q_k(t) &=& \sum_{r=1}^{N}
    (q_r (0) f_{r-k}(t) + p_r(0) g_{r-k}(t))
%    p_k(t) &=& \sum_{r=1}^{N} (
%    q_r (0) \partial_t {f}_{r-k}(t) + p_r(0) f_{r-k}(t)),
\end{eqnarray*}
and $p_k(t)=\partial_t q_k(t),$ where we have defined the two
functions
%\begin{eqnarray}
%   f_k(t)&=& \frac{1}{N}\sum_{l=1}^N
%   e^{2\pi i k l/N} \cos (\omega_l t),\\
%   g_k(t)&=& \frac{1}{N}\sum_{l=1}^N
%   e^{2\pi i k l/N} \sin (\omega_l t)/\omega_l.
%\end{eqnarray}
%\begin{eqnarray*}
%    f_k(t)&=& \sum_{l=1}^N
%    e^{\frac{2\pi i k l}{N}}
%    \frac{\cos (\omega_l t)}{N},\,
$    g_k(t)= \sum_{l=1}^N
    e^{2\pi i k l/N}
     \sin (\omega_l t)/(N\omega_l) $
     and
 $   f_k(t)=\partial_t g_k(t)$.
%\end{eqnarray*}
%\cite{Inf}
%$f_k(t)=\partial_t g_k(t)$
%
In this paper we are dealing with states that are Gaussian,
i.e., states whose characteristic function or Wigner function  is
a Gaussian. As such, it is completely characterized by the first
and second moments \cite{Eisert P 03}. The first moments will not
be directly relevant for our purposes. The second moments can be
arranged in the
symmetric $2N\times
2N$-covariance matrix
%\begin{equation}\nonumber
    $\Gamma_{R,S}= 2\text{Re} \langle
    (R- \langle R\rangle )
    (S- \langle S\rangle )
    \rangle,$
%\end{equation}
where $R$ and $S$ stand for the canonical operators
$q_1,\ldots,q_n$ and $p_1,\ldots,p_n$. At this point, we assume
that for times $t<0$, the  oscillators are not interacting and are
in the ground state.
This implies that
%\begin{eqnarray}
%    \langle q_{n} q_{m}\rangle &=&
%    \delta_{n,m} /2,\\
%    \langle p_{n} p_{m}\rangle &=&
%    \delta_{n,m} /2,\\
%    \langle q_{n} p_{m}\rangle &=&0,
%\end{eqnarray}
%\begin{eqnarray}\nonumber
    $\Gamma_{q_{n} q_{m}} =
    \Gamma_{p_{n} p_{m}} =
    \delta_{n,m}$, and
    $\Gamma_{q_{n} p_{m}} =0$,
%\end{eqnarray}
for $n,m=1,\ldots,N$.
% Well this is a non-interacting systems so that the entanglement
% is not only short range but actually zero!
%The ground state of this chain exhibits no
%long-range entanglement between pairs of oscillators.

 In the setting of this paper, we will assume for $t<0$ the
interaction is switched off and the system is in its ground state
and time-independent. At time $t=0$ the interaction is then
switched on instantaneously to ensure non-adiabaticity and
consequently the system is out of equilibrium and evolving in time
for $t>0$ according to the equations of motion for the second
moments given by
 \begin{eqnarray*}\nonumber
    \Gamma_{q_{n} q_{m}}(t) =
    (
    a_{n,m}(t) + d_{n,m}(t)
    )/2,\\
    \Gamma_{q_{n} p_{m}}(t)=
    (
    b_{n,m}(t) + e_{n,m}(t)
    )/2,\\
    \Gamma_{p_{n} p_{m}}(t)=
    (
    c_{n,m}(t) + a_{n,m}(t)
    )/2,
\end{eqnarray*}
where
$a_{n,m}= \sum_{k=1}^{N} f_{k-n} f_{k-m},$
%$b_{n,m}= \sum_{k=1}^{N} f_{k-n} \partial_t f_{k-m},$
$b_{n,m}=\partial_t a_{m,n}/2,$ $c_{n,m}= \sum_{k=1}^{N}
\partial_t f_{k-n} \partial_t f_{k-m},$, $d_{n,m} = \sum_{k=1}^{N}
g_{k-n} g_{k-m},$ and $e_{n,m}=\partial_t d_{n,m}/2$ .

Before we discuss in detail the non-adiabaticity requirement and
other idealizations as well as the physics behind this approach we
demonstrate the success of the approach.  We are now in the
position to study the entanglement of two very distant oscillators
when we ignore (trace out) all the others. The chain is
translationally invariant, and hence, a single oscillator, say
labeled $1$, can be singled out, and we may look at the degree of
entanglement as a function of time and discrete distance. We
quantify the degree of entanglement in terms of the logarithmic
negativity, defined as
%\begin{equation}
   $ E_N(\rho)= \log  \| \rho^{T_A} \|_1$
%\end{equation}
for states $\rho$, where $\rho^{T_A}$ is the partial transpose and
$\|.\|_1$ denotes the trace-norm.
%The negativity, which can
%be obtained from the covariance matrix $\Gamma$, is an
%entanglement monotone as it is non-increasing under local
%operations and classical communication and its
The logarithmic negativity is an upper bound for distillable
entanglement and has an interpretation of an asymptotic
preparation cost and bounds the distillable entanglement
\cite{Neg}.

Before we consider the entanglement created in this way, let us
first remind ourselves about the entanglement structure of the
ground state of the harmonic lattice Hamiltonian: there,  the
bi-partite entanglement between two distinguished oscillators is
only non-zero for nearest neighbors. Next-to-nearest neighbors
are already separable for all parameters, as are more distant
oscillators, even in case of an arbitrarily large correlation
length of the chain when approaching criticality, as has been
demonstrated in Ref.\ \cite{Chain}.

\begin{figure}[hbt]
\hspace*{-.4cm}\begin{minipage}{1.1\columnwidth}
   \includegraphics[width=4.1cm]{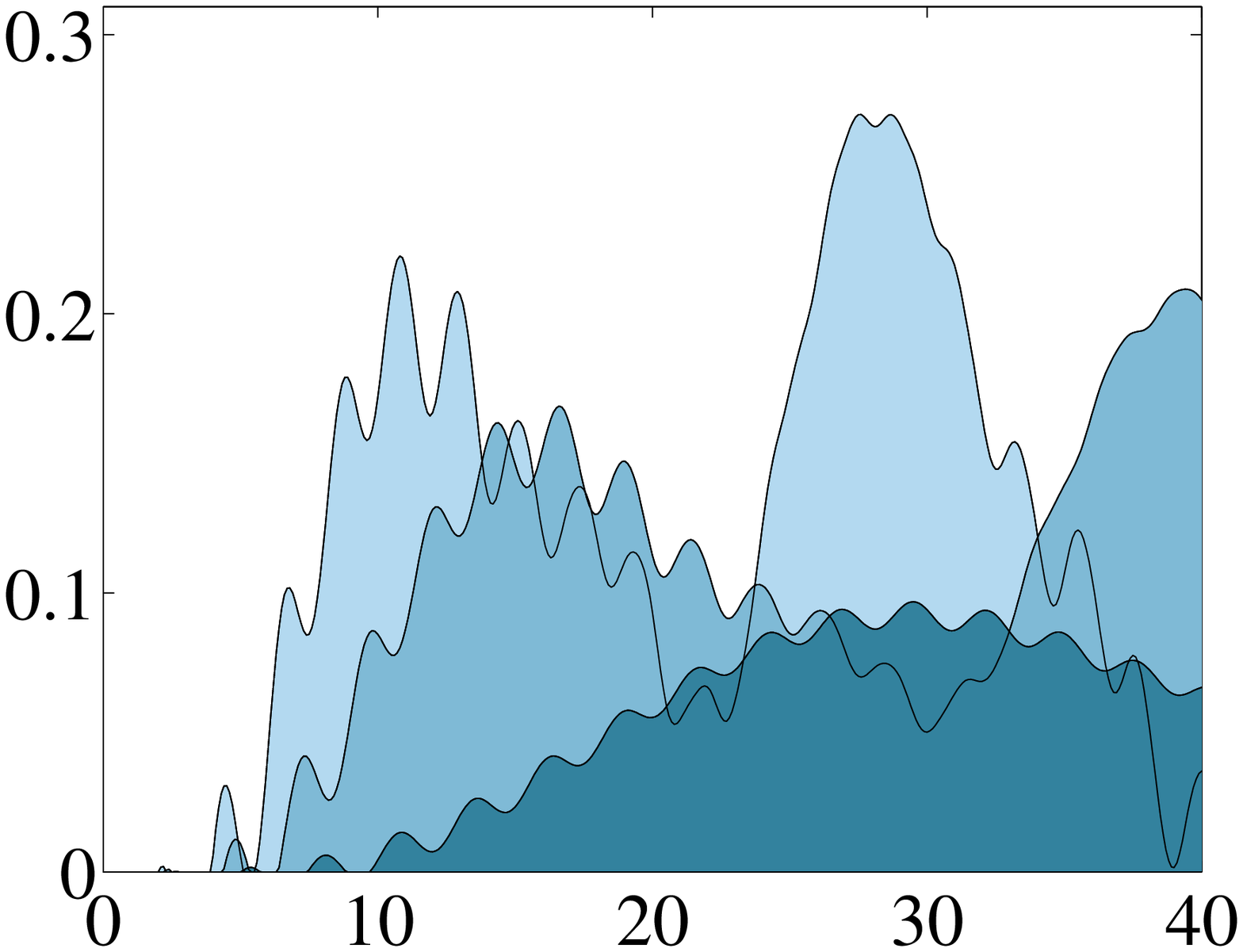}
   \includegraphics[width=4.1cm]{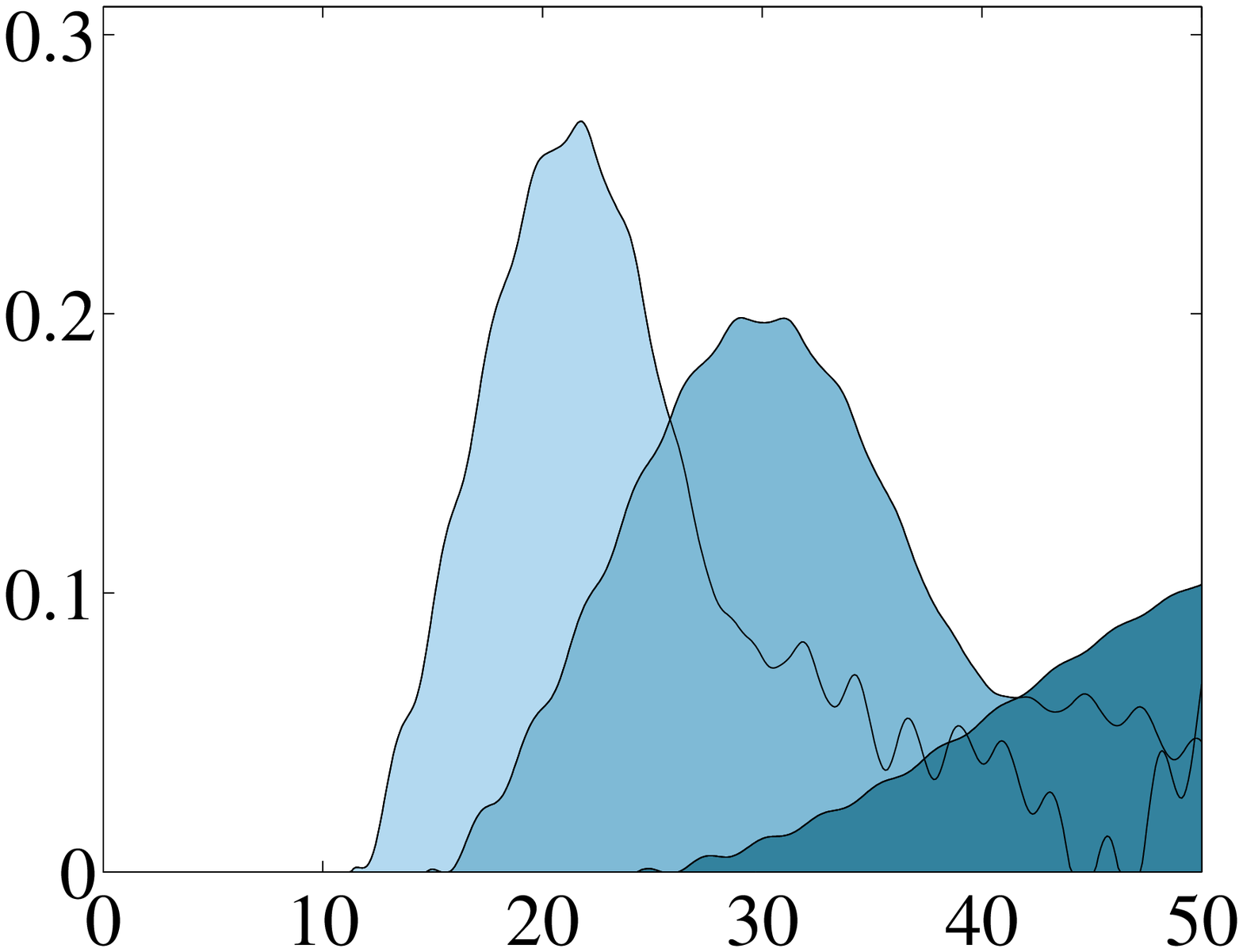}
\end{minipage}

\caption{The degree of entanglement as a function of time between
two oscillators  in a chain of length $8$ with periodic boundary
conditions (left) and open boundary conditions (right). On the
right hand side, the first and the last oscillator in the chain
are considered, on the left hand side the two diametrically
opposed oscillators. The values for $c$ are in the above units
$c=0.3$, $c=0.2$, and $c=0.1$ (depicted in light, medium, and dark
grey). }\label{f1}
\end{figure}

%
%
%Next-to-nearest neighbors are
%already separable for all parameters, as has been investigated in
%Ref.\ \cite{Chain}, let alone further separated oscillators,
%even in case of an arbitrarily large
%correlation length of the chain when approaching criticality.
This is very much in contrast to the situation encountered here:
Astonishingly indeed, we find that even very distant oscillators
become significantly entangled over time. This dependence is
depicted in Fig.\ \ref{f1} (for periodic boundary conditions
according to the above formalism, and numerically for open
boundary conditions). For a time interval $[0,t_{0})$, $t_{0}>0$,
the state of the oscillators with labels $1$ and $n$ is separable,
then, for $t>t_0$ it becomes entangled. This time $t_{0}$ is
approximately given by
%\begin{equation}\nonumber
    $t_{0} \approx n/(2\gamma\Omega),$
%\end{equation}
 There is what can be called a finite `speed of propagation'
of the quantum correlations, which is in fact closely related to
the speed of sound in this chain. The amount of entanglement
roughly falls off as $1/n$, but becomes strictly zero after a
finite distance. For $c=0.1$, for example, this happens for $n$
larger than $500$. This long-range nature of the entanglement is
remarkable indeed.

The central idea behind the method above is the well-known fact
that an instantaneous change in the potential of a single harmonic
oscillator in its ground state will  generally make its
state time dependent and squeezed. In the same way a change in the
coupling strength between oscillators drives the systems away from
equilibrium. In the course of the subsequent time evolution the
squeezing is then transformed into entanglement due to the
nearest-neighbor coupling. The origin for this is the fact that
the time evolution is described by a Hamiltonian quadratic in the
canonical coordinates and therefore has an effect analogous to
passive optical elements. It is well-known that a beamsplitter
which has squeezed states as an input will lead to entangled
outputs. This process happens continuously in the chain. Finally
this entanglement propagates, as every other excitation, through
the chain and can therefore lead to entanglement between distant
sites.

\begin{figure}[hbt]
\hspace*{-.4cm}
   \includegraphics[width=3.9cm]{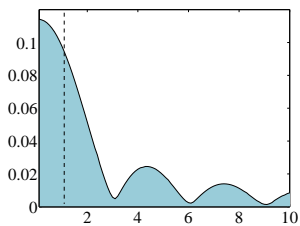}

\caption{Maximum degree of entanglement between the end points
as a function of the switching time $t'$
in the above units for an open chain of length $8$ and $c=0.1$. The vertical
dotted line represents the unit frequency of a free oscillator.}\label{AdiabaticFig}
  \end{figure}

In any realistic setting, this switching can not be instantaneous,
and an important question is how fast the switching process must
be in order to generate significant entanglement in the canonical
coordinates. Fig.\ \ref{AdiabaticFig} depicts the amount of
entanglement in the first maximum when the interaction strength is
linearly increased over a time interval $[0,t']$. We find that for
times $t'<1$, any non-zero switching time is unproblematic (with
very similar behavior found in longer chains). This is because the
change in coupling strength is faster than any eigenfrequency in
the system, preventing an adiabatic  following.
For very slow switching, $t'\gg1$, most
entanglement is lost because the system can adiabatically follow
the parameter change and remains approximately in the
 ground state.

Let us now turn to the discussion of a realisation in NEMS of such
an array. Presently, NEMS made from SiC have been manufactured
experimentally with frequencies around $1-10$ GHz, with spatial
dimensions of the order of $10$ nm \cite{Review}-\cite{Yang}.
Doubly clamped beams have the advantage of higher fundamental
frequencies with the same spatial dimensions. The $Q$-factors for
NEMS of these dimensions achieve values of  significantly more than $Q=10^{3}$
\cite{Higher,QFac}. Concerning the extent to which the ground
state can initially be reached, cooling of the oscillators to $10$
mK seems feasible \cite{Schwab2} using a helium dilution
refrigerator \cite{Roukes}, \cite{Schwab} (for the possibility of
the equivalent of laser cooling to the ground state, see Ref.\
\cite{Zoller}).

Needless to say, decoherence mechanisms cannot be entirely avoided
in a quantum system  so close to macroscopic dimensions.
After all, $Q$-factors describe nothing but the coupling strength
 to external degrees of freedom beyond our control. Most of the dissipation
 and decoherence is expected to be due to the coupling with the degrees
 of freedom of the substrate to which the resonator is connected.
 Let us now specify the decoherence model \cite{Dec}:
 In the setting described
here, we are not in the high temperature limit, but close to zero
temperature. Secondly, we do not have product initial conditions: in a
realistic setting, the chain and the environment are initially not
in a completely uncorrelated state, but rather in the
Gibbs state of the coupled joint system, and
then driven
away from equilibrium \cite{BroMo}.
We have hence modeled the decoherence process by appending local
heat baths consisting of a finite number $M$ of modes to each of
the oscillators with canonical coordinates
$q^{k}_{j},p^{k}_{j}$ for $k=1,\ldots,M$.
%$R_{j}=(q^{1}_{j},p^{1}_{j},\ldots, q^{M}_{j}, p^{M}_{j})$.
We choose a (discrete) Ohmic spectral density in
which case the Langevin equation for the Heisenberg picture
position becomes the one of classical Brownian motion in the
classical limit, i.e., the coupling is specified by the
interaction Hamiltonian
%\begin{equation}\nonumber
    $H_{j}= \zeta \tilde \omega_{j}  (
     q_{j} \otimes \sum_{i=1}^{M}q_{j}^{i} ),$
%\end{equation}
where $\tilde\omega_{j}= j\Lambda/M$, where $\Lambda>0$ is a cut-off
frequency. This Hamiltonian induces decoherence and dissipation,
and the number $\zeta>0$ has in our analysis been chosen in such a
manner that the energy dissipation rate reflects exactly the rate
$1/Q$ corresponding to the experimentally found $Q$-factors (see,
e.g., Refs.\ \cite{Harrington,Review}). With this value of
$\zeta$, the initial state before switching on the interaction is
then the Gibbs state of the canonical ensemble of the whole chain
including the appended heat baths. The resulting map
is nevertheless a Gaussian operation,
%, and will preserve the
%Gaussian character of the state,
such that it is sufficient to
know the second moments to specify entanglement properties. This
%decoherence
model grasps in the simplest possible manner the
various noise processes \cite{Cleland} in NEMS.

\begin{figure}

\hspace*{-.4cm}\begin{minipage}{1.1\columnwidth}
   \includegraphics[width=4.2cm]{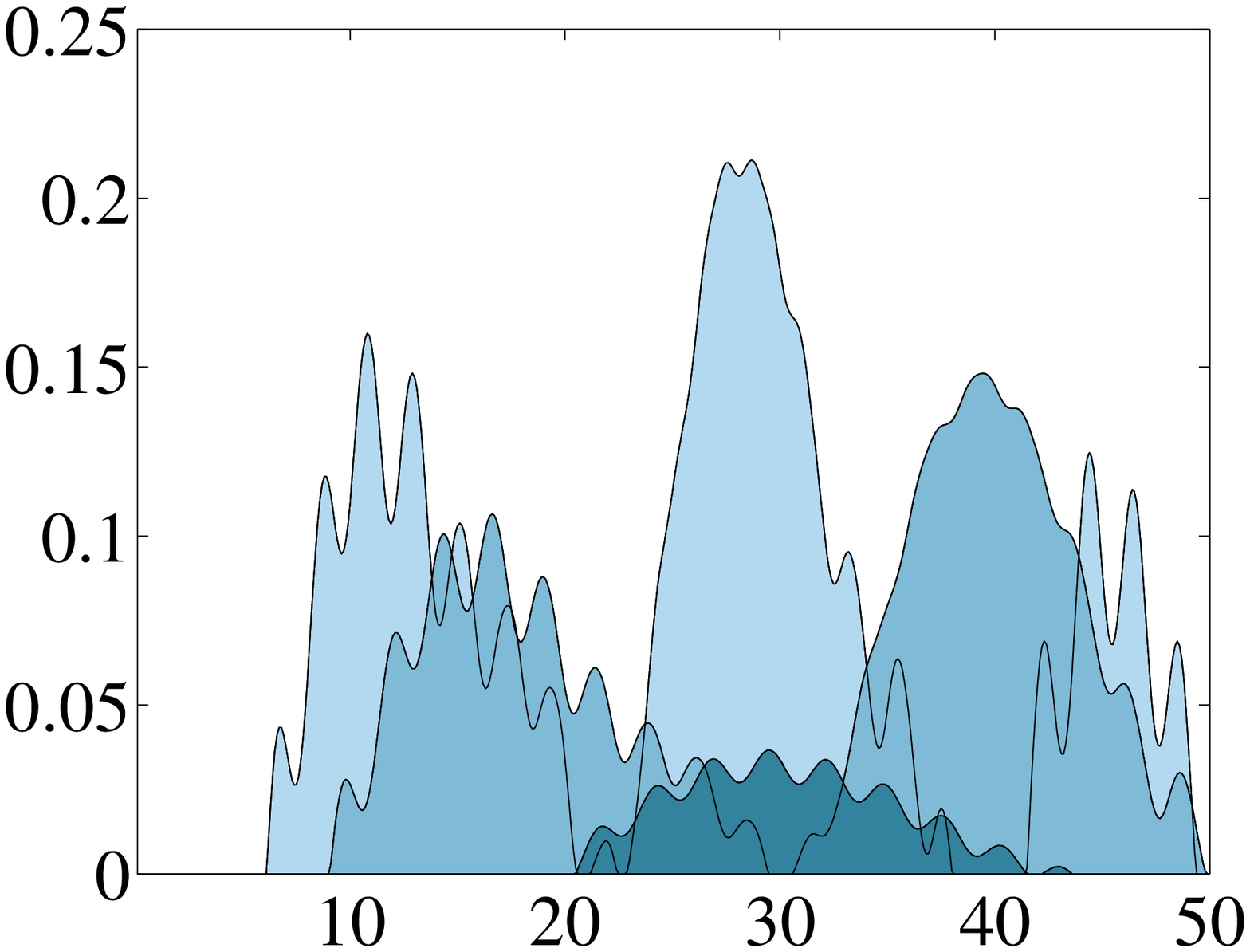}    \includegraphics[width=4.2cm]{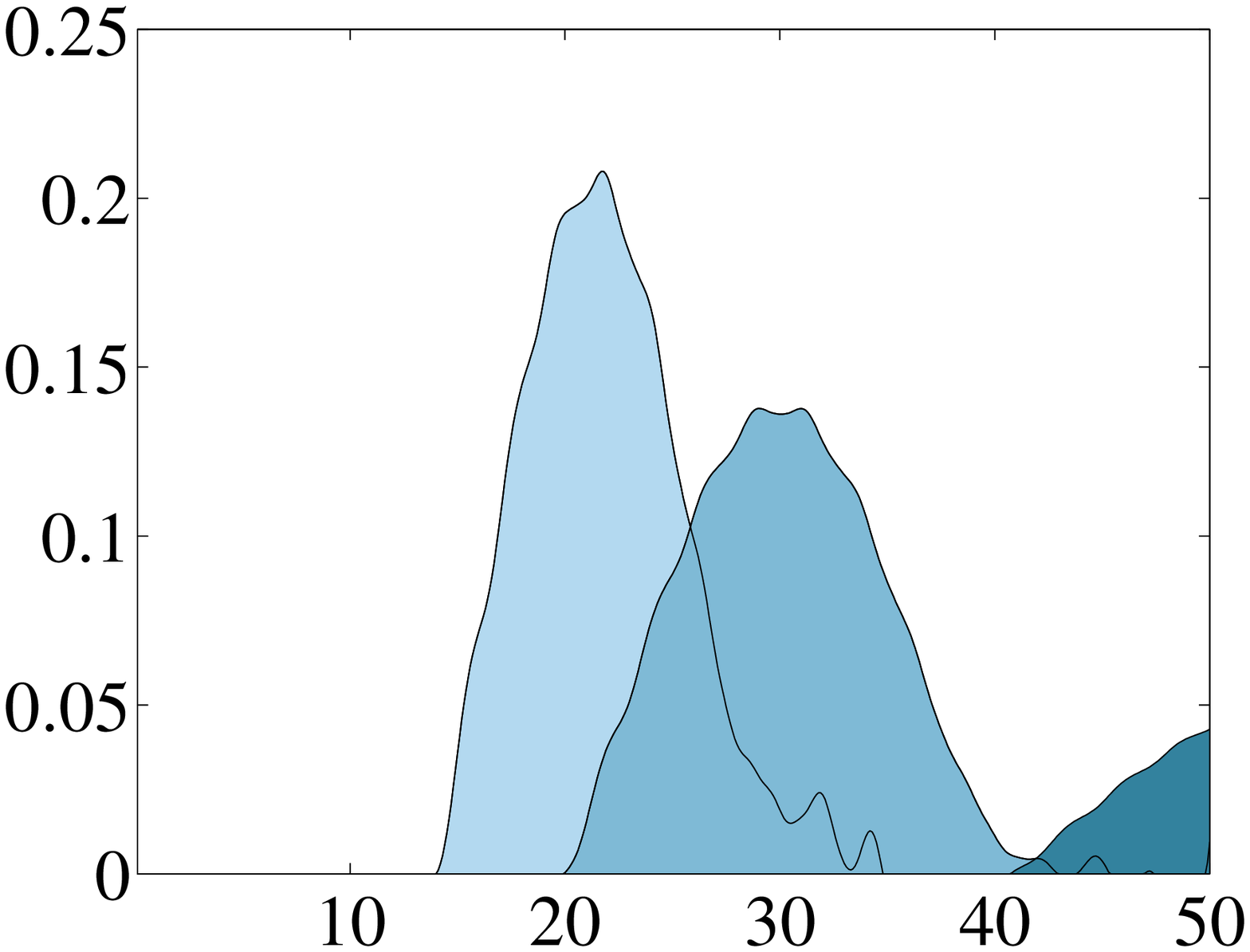}
\end{minipage}

\caption{The degree of entanglement under decoherence and for
non-zero temperature.
Shown is again the situation of a chain
of length $8$ for periodic (left) and open (right)
boundary conditions.
In this plot (up to rescaling of the
time axis, and a quantification of the coupling strength
$c=0.3$, $c=0.2$, and $c=0.1$ in terms of
the fundamental frequency),
values are chosen that correspond to
the $Q$-factor $Q=10^3$, fundamental frequency $5$ GHz, and
temperature of $10$ mK.
}\label{f2}
% Internal notes:
% For 20 oscillators a good value for Q=10 000 is
% zeta=8 x 10^-6 and delta=0.0001.
%
% For 10GHz and 10mK we have that beta=1/.13
\end{figure}

Fig.\ \ref{f2} depicts the behaviour of the degree of entanglement
for system parameters that are close to those used in actual
experimental settings. We see that the scheme is surprisingly
robust against noise processes and non-zero temperatures.
Comparably low $Q$-factors are not particularly harmful given the
large speed of propagation; yet too high temperatures, turn the
correlations into merely classical correlations. This effect is
evidently more harmful for longer chains. Notably, for two
oscillators, quite large values of the degree of entanglement can
be achieved. For example, for a two-oscillator system, with system
parameters as in Fig.\ \ref{f2}, the degree of entanglement as
quantified in terms of the log-negativity reaches values larger
than $0.6$ for $c=0.4$. Assuming the ability to cool to $10$ mK,
oscillators with fundamental frequencies of $2$ GHz would be
sufficient to generate entanglement. This would be the most
feasible starting point in such a scheme.

The most significant technological challenge in an experimental
realization of this scheme (and actually any scheme that involves
entanglement in the canonical coordinates of oscillators at the
nanoscale) is the actual detection of entanglement. We would
need to couple the two chosen oscillators to canonical coordinate
transducers whose output is proportional to position and momentum,
which is fed into an amplifier that produces a classical signal
\cite{Caves}. What has to be measured with very high sensitivity
are the second moments of the canonical coordinates $q_m$, $q_n$,
$p_m$, and $p_n$, i.e., covariance matrix elements. If not all
entries can be assessed, bounds of the type
%\begin{equation}
   $ E_{N}(\rho)\geq \max\left(0, -\log ((\langle (q_{n}-q_{m})^{2})\rangle + \langle
    (p_{n}+p_{m})^{2}\rangle)/2)\right)$
%\end{equation}
may be used to estimate the degree of entanglement.
If only a position transducer is available,
stroboscopic measurements may be employed where only two
measurements per cycle are performed (note that
$p_m(t=0)=q_m(t=\pi/4)$ and position and momentum are
interchanging roles with frequency $\omega$) \cite{Caves}.
Alternatively, continuous single-transducer measurements may be
performed which make use of only a position transducer and a
sinusoidally modulated output \cite{Caves}.
%It may also be easier
%to measure combinations $q_{m}-q_{n}$ and $q_{m}+q_{n}$ when
%ordering the oscillators in a ring structure.
This leaves us with
the problem of measuring position and momentum with great
accuracy: conventional optical transducers, as they can be
employed in MEMS, are not applicable in NEMS, but near-field
optical sensors or  piezoelectric detectors may be used
\cite{Review}. Refs.\ \cite{Ekinci,Measurement2} describe and make
use of a balanced electronic detection scheme of displacement. The
most promising to date appears to be a capacitive coupling of an
electrode placed on a resonator to the gate of a single-electron
transistor, as  studied theoretically in Ref.\ \cite{blencowe00}
and experimentally in Refs.\ \cite{Measurement,SchwabNew}.
The sensitivity
reached in such setups is rapidly increasing, and is presently
about a factor of $4.3$ away from the quantum limit of the
considered oscillator \cite{SchwabNew},
while this factor was still about a 100 a year ago,
and it is argued that with these
techniques, the quantum limit could well be reached in the not too far
future \cite{Measurement,blencowe00,SchwabNew,Schwab2}.

Finally, we would like to briefly mention that the chain of
mechanical oscillators may also be used in principle as a quantum
channel (compare also Ref.\ \cite{Bose}). If one feeds a half of a
highly entangled two-mode state into a harmonic chain with
nearest-neighbor interactions, then any oscillator of the chain
will at some time be entangled with the kept mode. The functional
behavior of the second moments as a function of time can be
approximated in terms of Bessel functions \cite{Other}, leading to
a time $t_{1}$ of the first arrival of entanglement at the $n$-th
oscillator of approximately (linear in $n$)
$
    t_{1}\approx 2 n/(\gamma \Omega e)$.

In this letter, we have presented an elementary method of
entangling mechanical oscillators on the nano-scale which are
located at macroscopically different locations at the ends of a
chain, without the need of addressing each of the oscillators in
the chain. We have introduced the suggested setup formally, and
have discussed issues of decoherence and measurement. As such, the
scheme is not yet a fully feasible scheme ready for
experimental implementation. Yet, it is the hope that this letter
can point towards significant next steps that could be taken when
further exploring the quantum domain with nano-electromechanical
devices.

This research was partly triggered by an inspiring talk given by
M.\ Roukes at CalTech in January 2003. J.E.\ would like to thank
J.\ Preskill and his
IQI group at CalTech for kind hospitality during a research
visit and S.B.\ would like to thank for a postdoctoral fellowship
at IQI. MBP is supported by a Royal Society Leverhulme Trust
Senior Research Fellowship. We would like to thank K.\ Schwab, M.\
Roukes, I.\ Wilson-Rae, P.\ Rabl, and C.\ Henkel for interesting communication. This work
has been supported by the EU (QUPRODIS), the DFG, the US Army
(DAAD 19-02-0161), and the EPSRC QIP-IRC.

%\end{multicols}
%
% Internal notes:
% For 20 oscillators a good value for Q=10 000 is
% zeta=8 x 10^-6 and delta=0.0001.
%
% For 10GHz and 10mK we have that beta=1/.13

%
% set(gca, 'FontName','Times')
% set(gca, 'FontSize',20)
% set(gca, 'XTickLabel', {2,4,6,8,10})
%
% area(erg,'FaceColor', [.6 .8 .9])


\begin{thebibliography}{99}
\bibitem{Higher}
The larger the spatial dimensions, the larger are the
$Q$-factors that have been achieved. The
logarithms of the best known $Q$-factors are
approximately linear in the logarithm of the
volume of the mechanical resonators \cite{Review}.
\bibitem{QFac}
The $Q$-factor specifies the dissipation of
energy to other modes in an uncontrolled manner.
It is the number of radians of oscillations necessary
for the energy to decrease by a factor of $1/e$.
\bibitem{BroMo}
Therefore, we have not modeled decoherence
by merely appending local terms reflecting decoherence in position
    $\partial_{t }\rho (t) = -i [H,\rho(t)]
    +
    \sum_{k=1}^{N} \epsilon_{k} [q_{k},[q_{k},\rho(t)]]$
to the generators of the dynamical map with suitable
$\epsilon_{k}>0$ (corresponding to the approximation of
the exact generator in Ref.\ \cite{Hu}
in the limit of very weak damping to the
extent of neglible friction, very high temperatures,
Ohmic spectral density, and factorizing initial conditions),
which is not necessarily appropriate, due
to the non-factorizing initial conditions.
\end{thebibliography}
\end{document}